\documentclass[12pt, draftclsnofoot, letterpaper, onecolumn, oneside]{./IEEEtran}

\usepackage{cite}
\usepackage{graphicx}
\usepackage{subfigure}
\usepackage{amsmath}
\usepackage{amssymb}
\usepackage{bm}
\usepackage{epsfig}
\usepackage{times}
\usepackage{color}
\usepackage{url}
\usepackage{array}
\usepackage{multirow}

\newtheorem{pavikl}{\textbf{Lemma}}

\newtheorem{pavikp}{\textbf{Proposition}}

\begin{document}

\title{Amplify-and-Forward in Wireless Relay Networks}%
\author{\IEEEauthorblockN{Samar Agnihotri, Sidharth Jaggi, and Minghua Chen}\\%
\IEEEauthorblockA{Department of Information Engineering, The Chinese University of Hong Kong}\\%
Email: \{samar, jaggi, minghua\}@ie.cuhk.edu.hk%
}

\maketitle

\begin{abstract}
A general class of wireless relay networks with a single source-destination pair is considered. Intermediate nodes in the network employ an amplify-and-forward scheme to relay their input signals. In this case the overall input-output channel from the source via the relays to the destination effectively behaves as an intersymbol interference channel with colored noise. Unlike previous work we formulate the problem of the maximum achievable rate in this setting as an optimization problem with no assumption on the network size, topology, and received signal-to-noise ratio. Previous work considered only scenarios wherein relays use all their power to amplify their received signals. We demonstrate that this may not always maximize the maximal achievable rate in amplify-and-forward relay networks. The proposed formulation allows us to not only recover known results on the performance of the amplify-and-forward schemes for some simple relay networks but also characterize the performance of more complex amplify-and-forward relay networks which cannot be addressed in a straightforward manner using existing approaches.

Using cut-set arguments, we derive simple upper bounds on the capacity of general wireless relay networks. Through various examples, we show that a large class of amplify-and-forward relay networks can achieve rates within a constant factor of these upper bounds asymptotically in network parameters.
\end{abstract}

\section{Introduction}
\label{sec:intro}
Since their introduction in \cite{104lanemanTseWornell} \textit{Amplify-and-Forward (AF)} relay schemes have been studied in the context of \textit{cooperative communication} \cite{107zhaoAdveLim, 107boradeZhengGallager}, estimating the capacity of relay networks \cite{105gastparVetterli, 107gomadamJafar, 109cuiHoKliewer}, and \textit{analog network coding} \cite{106zhangLiewLam, 107kattiMaricGoldsmith, 107kattiGollakottaKatabi, 110maricGoldsmithMedard, 110argyriouPandharipande}. For cooperative communication, AF schemes provide spatial diversity to fight against fading; for capacity estimation of relay networks, such schemes provide achievable lower bounds that are known to be optimal in some communication scenarios; and for analog network coding, given the broadcast nature of the wireless medium that allows the \textit{mixing} of the signals in the air, these schemes provide a communication strategy that achieves high throughput with low computational complexity at internal nodes. In this paper, we concern ourselves mostly with the capacity analysis of a general class of Gaussian AF relay networks. Extensions of our method and results to cooperative communication and analog network coding scenarios is part of our future work.

In previous work, while analyzing the performance of AF schemes in relay networks one or more of the following assumptions have been made: networks with a small number of nodes \cite{107kattiMaricGoldsmith, 110maricGoldsmithMedard}; networks with simple topologies \cite{105gastparVetterli, 107boradeZhengGallager, 107gomadamJafar, 107kattiMaricGoldsmith, 110maricGoldsmithMedard}; or relay operation in the high-SNR regime, \cite{110maricGoldsmithMedard}. However, for two reasons, we believe that it is important to characterize the performance of the AF schemes without such assumptions. First, we feel that for a scheme such as amplify-and-forward that allows one to exploit the broadcast nature of the wireless medium such assumptions on the size and topology may result in lower achievable performance than otherwise. Second, even in the low-SNR regimes amplify-and-forward can be capacity-achieving relay strategy in some scenarios, \cite{107gomadamJafar}. Therefore, a framework to address the performance of AF schemes in general wireless relay networks is desired.

However, one major issue with constructing such a framework is the following. In general wireless relay networks with AF relaying, the resulting input-output channel between the source and the destination is an intersymbol interference (ISI) channel ($\!\!$\cite{105gastparVetterli, 110maricGoldsmithMedard}) with colored noise. This is because both the source signal and the noise introduced at the relay nodes may reach the destination via multiple paths with differing delays. Without the assumptions above, this results in a formidable problem to analyze with the existing methods \cite{105gastparVetterli}.

Our main contribution is that we provide a framework to compute the maximum achievable rate with AF schemes for a class of general wireless relay networks, namely \textit{Gaussian relay networks}. This framework casts the problem of computing the maximum rate achievable with AF relay networks as an optimization problem. Our work shows that amplifying the received signal to the maximum possible value at intermediate nodes might result in sub-optimal end-to-end throughput. Also, we establish the generality of the proposed formulation by showing that it allows us to derive in a unified and simple manner not only the various existing results on the performance of simple AF relay networks but also new results for more complex networks that cannot be addressed in a straightforward manner with existing methods. We show through various examples that for a large class of relay networks the AF schemes can achieve rates within a constant factor of the cut-set upper-bounds on the capacity of general wireless relay networks.

The paper is organized as follows. In Section~\ref{sec:sysModel} we introduce the general class of Gaussian AF relay networks addressed in this paper. In Section~\ref{sec:achievableRate} we formulate the problem of maximum rate achievable via AF schemes in these networks. In Section~\ref{sec:computingInfoRate} we compute the rates achievable via AF schemes for two instances of such relay networks under various communication scenarios, and then in Section~\ref{sec:capacity} we discuss the asymptotic behavior of the gap between these rates and the corresponding upper bounds on the capacity of general wireless relay networks computed there. Section~\ref{sec:conclFW} concludes the paper with a summary.

\section{System Model}
\label{sec:sysModel}
\begin{figure}[!t]
\centering
\includegraphics[width=3.0in]{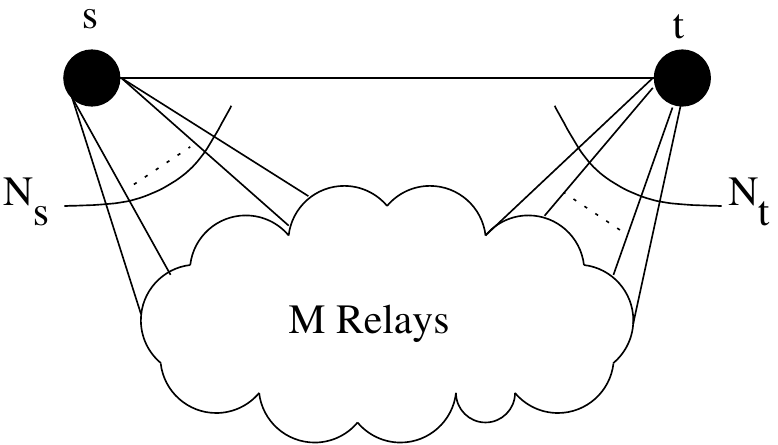}
\caption{A single source-single destination communication channel over general Gaussian relay network with $M$ relays.}
\label{fig:relayNw}
\end{figure}

Let us consider a $(M+2)$-node wireless relay network with source $s$ destination $t$ and $M$ relays as a directed graph $G=(V,E)$ with bidirectional links, as shown in Figure~\ref{fig:relayNw}. Each node in the network is assumed to have a single antenna. Let us assume that the degree of the source node is $N_s+1$, with it being connected to the destination node and a subset $S_s$ of the relay nodes, $N_s = |S_s|$. Similarly, let us assume that the degree of the destination node is $N_t+1$, with it being connected to the source node and a subset $S_t$ of the relay nodes, $N_t = |S_s|$. In general, $S_s \cup S_t \subseteq V \setminus \{s, t\}$.

At instant $n$, the channel output at node $i, i \in V \setminus \{s\}$, is
\begin{equation}
\label{eqn:channelOut}
y_i[n] = \sum_{j \in {\mathcal N}(i)} h_{ji} x_j[n] + z_i[n], \quad - \infty < n < \infty,
\end{equation}
where $x_j[n]$ is the channel input of the node $j$ in the neighbor set ${\mathcal N}(i)$ of node $i$. In \eqref{eqn:channelOut}, $h_{ji}$ is a real number representing the channel gain along the link from relay $j$ to relay $i$. It is assumed to be fixed (for example, as in a single realization of a fading process) and known throughout the network. Further, $\{z_i[n]\}$ is a sequence (in $n$) of independently and identically distributed (\textit{i.i.d.}) Gaussian random variables with zero mean and variance $\sigma^2, z_i[n] \sim {\cal N}(0, \sigma^2)$. We also assume that $z_i$ are independent of the input signal and of each other. The source symbols $x_s[n], - \infty < n < \infty$, are \textit{i.i.d.} Gaussian random variables with zero mean and variance $P_s$ that satisfies an average source power constraint, $x_s[n] \sim {\cal N}(0, P_s)$. We assume that the $i^{\textrm{th}}$ relay node's transmit power is constrained as:
\begin{equation}
\label{eqn:pwrConstraint}
E[x_i^2[n]] \le P_i, \quad - \infty < n < \infty
\end{equation}

In a general wireless relay network there may exists cycles. If a relay merely amplifies and forwards its received signal in such scenarios then it may be use a significant fraction of its power budget on forwarding the previously forwarded information. Therefore, motivated by some work on analog network coding, such as \cite{111youChenLiVucetic}, we propose the following relay operation to allow the relays to expend their transmit power in forwarding only the ``new'' information.

\textbf{\textit{Relay operation:}} We assume that each relay node maintains a buffer of signals it forwarded previously. Therefore, each relay node $i$, after receiving the channel output $y_i[n]$ at time instant $n$ executes the following series of steps:

\textit{Step 1:} Obtain the \textit{residual signal} $y_i^{\prime}[n]$ from its input $y_i[n]$ by subtracting the contributions of previously forwarded signals (if any) from $y_i[n]$.

\textit{Step 2:} Compute the power $P_{R,i}^{\prime}$ of the \textit{residual signal} $y_i^{\prime}[n]$.

\textit{Step 3:} At instant $n+1$ transmit the scaled version of the \textit{residual signal} $y_i^{\prime}[n]$ of its input at time instant $n$:
\begin{equation}
\label{eqn:AFdef}
x_i[n+1] = \beta_i y_i^{\prime}[n], \quad 0 \le \beta_i^2 \le \beta_{i,max}^2 = P_i/P_{R,i}^{\prime},
\end{equation}
where the $\beta_i$ is the scaling factor\footnote{Note that, in general, $\beta_i$ may depend on $i^{\textrm{th}}$ relay's past observations. $\beta_i[n] = f_{i,n}(Y_i[n-1], \ldots, Y_i[1])$. However, due to practical considerations, such as low-complexity operation, we do not consider such scenarios here.}. Let the network-wide amplification vector for the $M$ relay nodes be denoted as $\bm{\beta} = (\beta_1, \ldots, \beta_M)$.

\textit{Remark 1:} One of the major advantages of this relay operation is that by subtracting the previously forwarded signal from its input, a relay node expends its power in forwarding only the ``new'' information. In a general wireless relay network there may exists cycles. In such scenarios if a relay merely forwards its received signal then it may result in the relay forwarding the scaled version of the linear combination of two or more of its previously forwarded signals. This may provide higher achievable rate as in \cite{104zahediMohseniGamal}.

Using \eqref{eqn:channelOut} and \eqref{eqn:AFdef}, the input-output channel between the source and destination can be written as an intersymbol interference (ISI) channel that at instant $n$ is given by
\begin{eqnarray}
y_t[n] &=&  h_{st} x_s[n] + z_t[n] \label{eqn:ISIchnl} \\
 &+&  \sum_{d=1}^{D^s}\bigg[\sum_{(i_1, \ldots, i_{d}) \in K_{d}} h_{si_1}\beta_{i_1}h_{i_1 i_2} \ldots h_{i_{d-1}i_{d}}\beta_{i_{d}}h_{i_{d} t}\bigg] x_s[n-d] \nonumber \\
 &+&  \sum_{d=1}^{D^1}\bigg[\sum_{(i_1, \ldots, i_{d}) \in K_{1,d}} \beta_1 h_{1 i_1} \ldots h_{i_{d-1}i_{d}}\beta_{i_{d}}h_{i_{d} t}\bigg] z_1[n-d] \nonumber \\
 && \vdots \nonumber \\
 &+&  \sum_{d=1}^{D^M}\bigg[\sum_{(i_1, \ldots, i_{d}) \in K_{M,d}} \beta_M h_{M i_1} \ldots h_{i_{d-1}i_{d}}\beta_{i_{d}}h_{i_{d} t}\bigg] z_M[n-d], \nonumber
\end{eqnarray}
where $K_d, 1 \le d \le D^s$, is the set of $d$-tuples of node indices corresponding to all paths from the source to the destination with path delay $d$ and $D^s$ is the length of the longest such path. Note that along such paths $D^s \le M$. Similarly, $K_{m,d}, 1 \le m \le M, 1 \le d \le D^m$, is the set of $d$-tuples of node indices corresponding to all paths from the $m^{\textrm{th}}$ relay to the destination with path delay $d$, $D^m$ is the length of the longest such path from $m^{\textrm{th}}$ relay to the destination. It should be noted that $\max(D^1, \ldots, D^M) = D^s-1$.

Let us introduce \textit{modified} channel gains as follows. For all the paths between the source $s$ and the destination $t$:
\begin{eqnarray}
&& h_0 = h_{st} \label{eqn:modChnlParams}\\
&& h_d = \sum_{(i_1, \ldots, i_{d}) \in K_{d}} h_{si_1}\beta_{i_1}h_{i_1 i_2} \ldots h_{i_{d-1}i_{d}}\beta_{i_{d}}h_{i_{d} t}, 1 \le d \le D^s \nonumber
\end{eqnarray}
For all the paths between the $m^{\textrm{th}}$-relay, $1 \le m \le M$, and the destination $t$:
\begin{eqnarray}
&& h_{m,0} = 0, \label{eqn:modChnlParams2} \\
&& h_{m,d} = \sum_{(i_1, \ldots, i_{d}) \in K_{m,d}} \beta_m h_{m i_1} \ldots \nonumber h_{i_{d-1}i_{d}}\beta_{i_{d}}h_{i_{d} t}, 1 \le d \le D^m \nonumber
\end{eqnarray}

\textit{Remark 2:} Note that though there may be exponentially large number of paths between the source and the destination as well as between a relay and the destination, the modified channel gains for all such paths as in \eqref{eqn:modChnlParams} and \eqref{eqn:modChnlParams2} can be efficiently computed using the line-graphs \cite{103koetterMedard}.

In terms of these modified channel parameters, the source-destination ISI channel in \eqref{eqn:ISIchnl} can be written as:
\begin{equation}
\label{eqn:ISIchnlmod}
y_t[n] = \sum_{j=0}^{D^s} h_j x_s[n-j] + \sum_{j=0}^{D^1} h_{1,j} z_1[n-j] + \ldots + \sum_{j=0}^{D^M} h_{M,j} z_M[n-j] + z_t[n]
\end{equation}

Before proceeding further, let us introduce two special cases of the general class of relay networks introduced in the beginning of this section. In the rest of this paper, we illustrate various concepts and derive some results using these two special networks.

\begin{figure}[!t]
\centering
\includegraphics[width=3.0in]{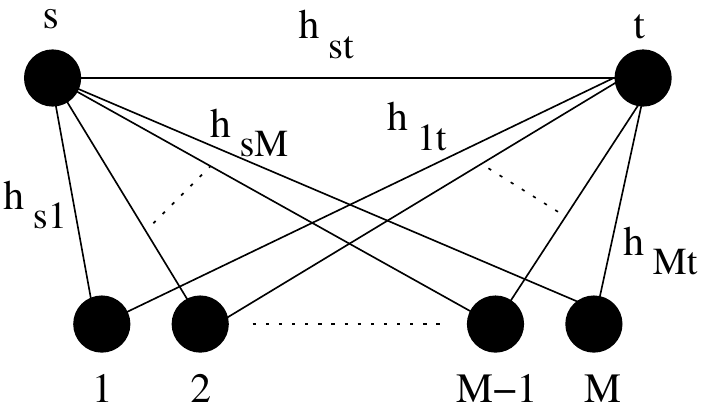}
\caption{Type A Relay Network.}
\label{fig:relayNwType1}
\end{figure}

\begin{figure}[!t]
\centering
\includegraphics[width=3.0in]{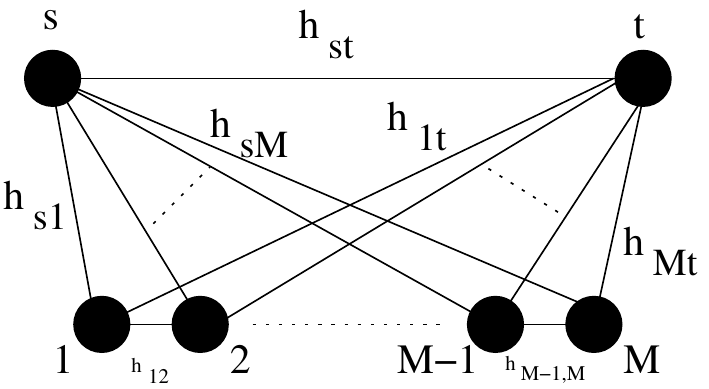}
\caption{Type B Relay Network.}
\label{fig:relayNwType2}
\end{figure}

\textit{Type A Network:} For one source-destination pair and $M$ relays, Type A network is defined as: $G= (V,E)$, where $V=\{s, t, 1, \ldots, M\}$ and $E = \{(s,t), (s,i), (i,t): i \in \{1, \ldots, M\} \}$. In other words, the source node shares an edge with the destination node and each of $M$ relay nodes. Similarly, the destination node shares an edge with the source node and each of $M$ relay nodes. However, no pair of relay nodes share an edge between themselves. Figure~\ref{fig:relayNwType1} illustrates an instance of Type A networks.

\textit{Type B Network:} For one source-destination pair and $M$ relays, Type B networks are defined as: $G= (V,E)$, where $V=\{s, t, 1, \ldots, M\}$ and $E = \{(s,t), (s,i), (i,t), (j, j+1): i \in \{1, \ldots, M\}, j \in \{1, \ldots, M-1\} \}$. In other words, as in Type-A networks, the source node shares an edge with the destination node and each of $M$ relay nodes. Similarly, the destination node shares an edge with the source node and each of $M$ relay nodes. However, $j^{\textrm{th}}$ relay shares an edge with $j+1^{\textrm{st}}$ node, for all $j \in \{1, \ldots, M-1\}$ . Figure~\ref{fig:relayNwType2} illustrates an instance of Type B networks.

\section{Achievable rates for the source-destination ISI channel in general AF Relay Networks}
\label{sec:achievableRate}
We first derive the expression for the achievable rate for the source-destination channel in \eqref{eqn:ISIchnlmod} for a given amplification-vector $\bm{\beta}$ and then formulate the problem of maximizing the achievable rate over the domain of feasible $\bm{\beta}$.

\begin{pavikl}
\label{lemma:infoRate}
For given length-$M$ vector $\bm{\beta}$, the achievable rate for the channel in \eqref{eqn:ISIchnlmod} with \textit{i.i.d.} Gaussian input is:
\begin{equation}
\label{eqn:infoRateFin}
I(P_s, \bm{\beta}) = \frac{1}{2 \pi} \int_{0}^{\pi} \log\bigg[1 + \frac{P_s}{\sigma^2} \frac{|H(\lambda)|^2}{1 + \sum_{m=1}^M |H_{m}(\lambda)|^2}\bigg] d\lambda, 
\end{equation}
where
\begin{equation}
\label{eqn:trnsFn}
H(\lambda) = \sum_{j=0}^{D^s} h_j e^{-ij\lambda}, \: H_m(\lambda) = \sum_{j=0}^{D^m} h_{m,j} e^{-ij\lambda}, \: i = \sqrt{-1} 
\end{equation}
\end{pavikl}
\begin{IEEEproof}
In \cite{088hirtMassey} a \textit{Discrete Fourier Transform (DFT)} based formalism is developed to compute the capacity of Gaussian channel with source intersymbol interference (ISI). We compute the maximum achievable rate for the channel in \eqref{eqn:ISIchnlmod} for a given amplification-vector $\bm{\beta}$ by generalizing this formalism to also include the ISI channel for the Gaussian noise at each relay node resulting in colored Gaussian noise at the destination\footnote{Note that the capacity of discrete-time Gaussian ISI channel was known prior to \cite{088hirtMassey}, for example in \cite{070tsybakov, 073tomsBerger, 074brandenburgWyner}. However, the analysis in these papers, based on the asymptotic properties of Toeplitz forms \cite{058grenanderSzego, 072gray}, is not easily amenable to derive the capacity results of the ISI channel with colored Gaussian noise. Therefore, like \cite{101goldsmithEffors, 109choudhuriMitra}, we too use more accessible \textit{DFT} based formalism developed in \cite{088hirtMassey} to compute the maximum achievable rate for the ISI channel with colored Gaussian noise in \eqref{eqn:ISIchnlmod}.}. The details of the proof are in the Appendix~\ref{appndx:proofInfoRate}.
\end{IEEEproof}

\textit{Remark 3:} The derivation of an expression for $I(P_s, \bm{\beta})$ with jointly Gaussian inputs is similar to the proof of Lemma~\ref{lemma:infoRate}. However as such an expression does not aid in the presentation of our ideas we do not discuss it in this paper.

For a given network-wide amplification vector $\bm{\beta}$, the achievable information rate is given by $I(P_s, \bm{\beta})$. Therefore the maximum information-rate $I_{AF}(P_s)$ achievable in an AF relay network with \textit{i.i.d.} Gaussian input is defined as the maximum of $I(P_s, \bm{\beta})$ over all feasible $\bm{\beta}$, subject to per relay-node amplification constraint \eqref{eqn:AFdef}. In other words:
\begin{equation}
\label{eqn:maxAFrate}
\hspace{-0.85in}\mbox{(P1): } \qquad I_{AF}(P_s) \stackrel{def}{=} \max_{\bm{\beta}:0 \le \beta_i^2 \le \beta_{i, max}^2} I(P_s, \bm{\beta})
\end{equation}
Substituting for $H(\lambda)$ and $H_m(\lambda)$ from \eqref{eqn:trnsFn} in \eqref{eqn:infoRateFin}, we can rewrite problem \eqref{eqn:maxAFrate} equivalently as:
\begin{eqnarray}
\mbox{(P2): } I_{AF}(P_s)  &=&  \max_{\bm{\beta}:0 \le \beta_i^2 \le \beta_{i, max}^2} I(P_s, \bm{\beta}) \label{eqn:maxAFrateExpanded} \\
I(P_s, \bm{\beta})  &=&  \frac{1}{2 \pi} \int_{0}^{\pi} \log\bigg[1 + \frac{P_s}{\sigma^2} \frac{\sum_{i=0}^{D^s} A_{i} \cos(i \lambda)}{\sum_{i=0}^{D^n} B_{i} \cos(i \lambda)}\bigg] d\lambda \nonumber
\end{eqnarray}
The coefficients $A_i, 1 \le i \le D^s$, and $B_i, 1 \le i \le D^n$, are defined in Appendix~\ref{appndx:explicitExprssns}.

The formulation of the problem P2 is illustrated in Appendix~\ref{appndx:probFormulnIllustrn} for the Type A and Type B relay networks introduced in the previous section.

In general AF relay networks the simultaneous relay transmissions may interfere and if the relays always amplify the received signals to the maximum possible then it may result in sub-optimal end-to-end throughput. Therefore the scaling factor for each relay must be optimally chosen to maximize the achievable rate. This is emphasized by P1 and its significance is illustrated by the following example.

\begin{figure}[!t]
\centering
\includegraphics[width=3.0in]{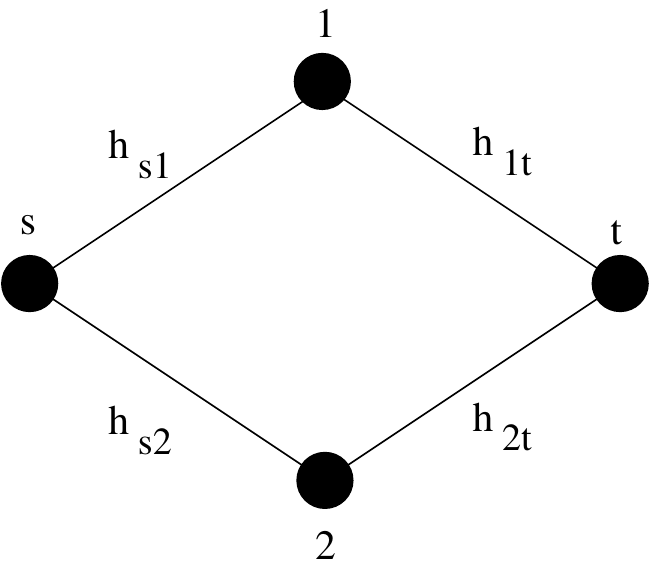}
\caption{Diamond network.}
\label{fig:diamond}
\end{figure}

\begin{figure}[!t]
\centering
\includegraphics[width=3.0in]{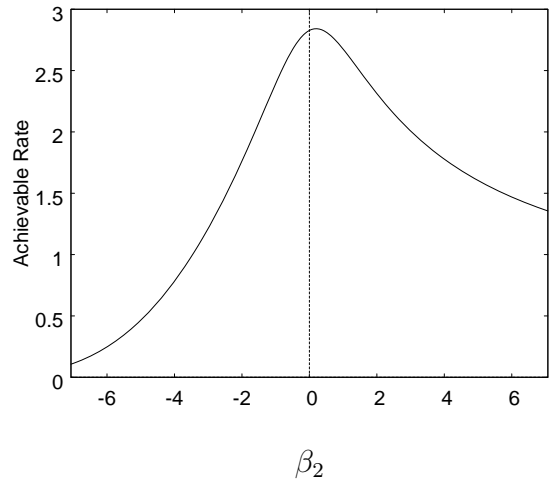}
\begin{center}
\vspace{-0.25in}
\hspace{0.25in}$\beta_2$
\end{center}
\caption{The achievable rate for the Example 1 when $\beta_1 = \beta_{1,max}$ and $\beta_2$ lies in $[-\beta_{2,max}, \beta_{2,max}]$.}
\label{fig:betaPlot}
\end{figure}

\textbf{Example 1:} Let us consider the relay network in Figure~\ref{fig:diamond}. Let $h_{s1} = 1, h_{s2} = 0.1, h_{1t} = h_{2t} = 1$. Let $P_s = P_1 = P_2 = 10$ and noise variance $\sigma^2 = 0.1$ at each node. Therefore, we have
\begin{eqnarray*}
\beta_{1,max}^2 = \frac{P_1}{h_{s1}^2 P_s + \sigma^2} = 0.99 \\
\beta_{2,max}^2 = \frac{P_2}{h_{s2}^2 P_s + \sigma^2} = 50.0
\end{eqnarray*}

In this case from \eqref{eqn:maxAFrate}, we have the following rate maximization problem:
\begin{equation*}
I_{AF} = \max_{\beta_1, \beta_2}\frac{1}{2} \log\bigg[1 + 100 \frac{(\beta_1 + 0.1 \beta_2)^2}{1 + \beta_1^2 + \beta_2^2}\bigg]
\end{equation*}
subject to constraints $0 \le \beta_1^2 \le \beta_{1,max}^2$ and $0 \le \beta_2^2 \le \beta_{2,max}^2$.

The optimal solution of this problem is $(\beta_1 = 0.995, \beta_2 = 0.225)$. The objective function is plotted in the Figure~\ref{fig:betaPlot} for $\beta_1 = 0.995$. Therefore, it follows that in this case $\beta_2 = \beta_{2,max}$ is not the optimal amplification factor. \hspace*{\fill}~\IEEEQEDclosed\par

With this observation and the definition of the relay operation given in the previous section, it is appropriate to call the forwarding scheme proposed in this paper as \textit{subtract-scale-and-forward}.

\textit{Remark 4 (the computational complexity of the problem P2):} Writing the objective function of the problem P2 as sum of ratios, as in \eqref{eqn:appndxInfoRate}, we can show from \cite{088avrielDiewertSchaible} that each of these ratios is neither quasiconcave\footnote{A function $f:S \rightarrow R$ defined on a convex subset S of real vector space is quasiconcave if whenever $x, y \in S$ and $\lambda \in [0,1]$, then $f(\lambda x + (1-\lambda)y) \ge \min(f(x), f(y))$, \cite{088avrielDiewertSchaible}.} (therefore, more than one local optimum may exist) nor quasiconvex (therefore, the globally optimal solution may not exist at an extreme point or on the boundary of the domain of optimization). The sum of such non-quasiconcave and non-quasiconvex ratios is also non-quasiconcave and non-quasiconvex. Therefore, the problem P2 is a global optimization problem, specifically it belongs to a subset of global optimization problems, called \textit{sum-of-ratios} problems, which are supposed be hard, in general, \cite{103schaibleShi}. Further, in general, $\beta_{i,max}, 1 \le i \le M$, depends on $\bm{\beta}_{-i} \stackrel{def}{=} (\beta_1, \ldots, \beta_{i-1}, \beta_{i+1}, \ldots, \beta_M)$. This dependence of $\beta_{i,max}$ on $\bm{\beta}_{-i}$ makes the constraint set in \eqref{eqn:maxAFrateExpanded} non-convex for all but Type A relay networks. Based on these arguments, we conjecture the problem P2 to be computationally hard. However, at present we do not have a formal computational complexity-theoretic proof and constructing such a proof is the part of our future work.

\textit{Remark 5 (the approximation schemes to solve the problem P2):} The lack of the exact complexity classification of the optimization problem in \eqref{eqn:maxAFrateExpanded} notwithstanding, we envision solving this problem to be an important step towards characterizing the maximum rates achievable with various other relaying schemes in general wireless relay networks as such schemes also result in problem formulations similar to \eqref{eqn:maxAFrateExpanded}. In \cite{111agnihotriChenJaggi}, we concern ourselves with developing a unified framework to efficiently approximate the optimization problems such as \eqref{eqn:maxAFrateExpanded}.

In the following we show that the problem formulation above not only allows us to reproduce existing results on the achievable rates for some special classes of amplify-and-forward wireless relay networks in a simple manner, but also allows us to compute the achievable rates for much broader class of amplify-and-forward wireless relay networks which could not be hitherto addressed with existing methods.

\section{Approximating the Maximum Achievable Rate $I_{AF}(P_s)$ for Amplify-and-Forward Relay Networks}
\label{sec:computingInfoRate}
Let us consider the problem P2 when the relays operate instantaneously as in \cite{107kattiMaricGoldsmith}, that is, the relays amplify-and-forward their input signals without delay. Therefore, for relay node $i, 1 \le i \le M$, we have $x_i[n] = \beta_i y_i^{\prime}[n]$. Note the possible system instability resulting from this assumption is avoided by the relay-operation (buffering and subtracting of the previously forwarded signals) as given in the Section~\ref{sec:sysModel}. With this assumption, P2 reduces to (after setting $\lambda = 0$ in \eqref{eqn:maxAFrateExpanded} and then integrating):
\begin{equation}
\label{eqn:maxAFrateNoDelatRel}
\mbox{(P3): } I_{AF}(P_s) = \max_{\bm{\beta}:0 \le \beta_i^2 \le \beta_{i, max}^2} \frac{1}{2} \log\bigg[1 + \frac{P_s}{\sigma^2} \frac{\sum_{i=0}^{D^s} A_{i}}{\sum_{i=0}^{D^n} B_{i}}\bigg] 
\end{equation}
The problem P3 can also be derived directly from the channel model in \eqref{eqn:ISIchnlmod}, however we do not discuss that derivation here for the sake of brevity.

Given the particular form of functional dependence of $A_i$s and $B_i$s on the components of the amplification vector $\bm{\beta}$ (as follows from \eqref{eqn:ISIchnl}, \eqref{eqn:modChnlParams}, \eqref{eqn:appndxBsig2}, and \eqref{eqn:appndxBnoiseSum1}) and dependence of $\beta_i$s on each other, the problem P3 is a \textit{Geometric Program (GP)} \cite{107boydKimVandenberghe}, where both, the objective function as well as the constraints are expressed in terms of \textit{posynomials}. However, minimizing or upper bounding (equivalently, maximizing or lower bounding) a ratio of two posynomials is a non-convex problem that is intrinsically intractable \cite{105chiang, 107chiangTanPalomar}\footnote{Note that in \cite{105chiang, 107chiangTanPalomar} the objective function can be written as an inverted posynomial in high-SINR regime. Therefore the problem is efficiently solvable using GP methods. It is only in low to moderate SINR regimes where the objective function cannot be so written, the problem is NP-hard. However, the objective function in \eqref{eqn:maxAFrateNoDelatRel} involves the ratio of posynomials for all SNR values. Therefore, in general, the problem is hard irrespective of SNR value.}. Therefore, for a general relay network, it is not possible to exactly solve or lower bound the problem P3 in a computationally efficient manner. However, unlike the problem P2 in \eqref{eqn:maxAFrateExpanded}, for the problem P3, efficient approximation schemes exist that solve such problems iteratively by solving a series of GPs, as discussed in \cite[Section 3.3]{105chiang}. Also, for some specific relay networks, under some assumptions, we can efficiently compute the lower bounds, as we show next.

Let us consider the problem P3 when the relay nodes are constrained to use the same amplification-factor, that is, $\beta_i = \beta$, for all $1 \le i \le M$. In the practical setting, this assumption considerably simplifies the system-design with $\beta$ set to one particular value for all relay nodes. Then, (P3) reduces to
\begin{eqnarray}
\mbox{(P4): } \hspace{0.5in} I_{AF}(P_s) = \max_{0 \le \beta^2 \le \beta_{max}^2} I_{AF}(P_s, \beta) && \label{eqn:maxAFrateEqBeta} \\
I_{AF}(P_s, \beta) = \frac{1}{2} \log\bigg[1 + \frac{P_s}{\sigma^2} \frac{\sum_{i=0}^{D^s} A_{i}}{\sum_{i=0}^{D^n} B_{i}}\bigg] && \label{eqn:maxAFrateFnOBeta}
\end{eqnarray}

Note that the solution of (P3) cannot be smaller than the solution of (P4) because the set of feasible $\bm{\beta}:\beta_i = \beta$, for (P4) is a subset of the set of feasible $\bm{\beta}$ for (P3).

\textbf{Note 1:} In general, $\beta_{i,max}^2 \sim P_i$. Therefore $\sum \beta_{i,max}^2 \sim \sum P_i$. However, as we are considering the case of equal $\beta$, $\beta_{i,max} = \beta_{max}$, so we have $M \beta_{max}^2 \sim \sum P_i$ or $\beta_{max}^2 \sim M^{-1} \sum P_i$.

Now we discuss solving the problem P4 for Type-A and Type-B relay networks introduced in Section~\ref{sec:sysModel}, in different communication scenarios. The proofs of various propositions and lemmas are provided in the Appendix~\ref{appndx:proofsOLemmasProps}.

\subsection{Type-A Relay Network}
\label{subsec:typeA}
Let us first consider Type-A relay network as in Figure~\ref{fig:relayNwType1}. For such networks, we solve the problem P4 in the following two scenarios.

\textbf{\textit{Scenario 1 (No attenuation network):}} Let us assume that there is no attenuation along any link in the network, that is, $h_{st} = h_{si} = h_{it} = 1$ for all $1 \le i \le M$. The problem P4 in this case is:
\begin{pavikp}
\label{prop:exa1rate}
\begin{equation*}
I_{AF}(P_s) = \max_{0 \le \beta^2 \le \beta_{max}^2} \frac{1}{2} \log\bigg[1 + \frac{P_s}{\sigma^2} \frac{(1 + M \beta)^2}{1 + M \beta^2}\bigg]
\end{equation*}
\end{pavikp}

\begin{pavikl}
\label{lemma:typeAnoAttenuation}
$I_{AF}(P_s)$ attains its global maximum at $\beta_{opt}=1$.
\end{pavikl}

Now let us consider two particular ways in which $\beta_{max}$ varies with network size.

\textit{Scenario 1, Case A (Increasing relay power):} For the given network with $M$ relay nodes, let us assume that the sum power of the relay nodes is constrained as follows:
\begin{equation*}
\sum_{m=1}^M E[X_m^2] \le \sum_{m=1}^M P_i \le M^{u+1} Q, u > 0, Q = constant
\end{equation*}
So $\beta_{max}^2 = M^u Q$. From Lemma~\ref{lemma:typeAnoAttenuation}, we have for $M \rightarrow \infty$.
\begin{equation}
\label{eqn:upRelPwr}
I_{AF}(P_s) = \left\{
                    \begin{array}{ll}
                    \frac{1}{2} \log[1 + \frac{P_s}{\sigma^2} (1 + M)], \mbox{ if } \beta_{max} \ge 1,\\
                    \frac{1}{2} \log[1 + \frac{P_s}{\sigma^2} (M + \frac{2}{\beta_{max}})], \mbox{ otherwise}
                    \end{array}
            \right.
\end{equation}

\textit{Scenario 1, Case B (Constant total relay power):} Let us consider the case where the sum power of relay nodes is fixed irrespective of the number of relay nodes, that is $\sum_{m=1}^M P_i \le Q, Q = constant$. Therefore, we set $\beta_{max}^2 = \frac{Q}{M}$. As $M \rightarrow \infty$, for sufficiently large $M$, $\beta_{max} < \beta_{opt} = 1$. Therefore, from Lemma~\ref{lemma:typeAnoAttenuation}, $\beta=\beta_{max}$ maximizes the achievable rate and we have for $M \rightarrow \infty$
\begin{equation}
\label{eqn:constRelPwrType1}
I_{AF}(P_s) = \frac{1}{2} \log\bigg[1 + \frac{P_s}{\sigma^2} \frac{Q}{1+Q}M\bigg]
\end{equation}

\textbf{\textit{Scenario 2 (Bounded channel gains):}} In Scenario 1, we considered no attenuation relay networks. Now, let us consider the scenario where the channel gains are arbitrary, but strictly bounded, $0 < h_{st}, h_{si}, h_{it} < \infty$, $1 \le i \le M$. The problem P4 in this case is:
\begin{pavikp}
\label{prop:exa1ratebddgains}
\begin{equation*}
I_{AF}(P_s) = \max_{0 \le \beta^2 \le \beta_{max}^2}  \frac{1}{2}  \log  \bigg[ 1  +  \frac{P_s}{\sigma^2} \frac{(h_{st} + \beta \sum_{i=1}^{M} h_{si} h_{it})^2}{1 + \beta^2 \sum_{i=1}^M h_{it}^2}\bigg]
\end{equation*}
\end{pavikp}

\begin{pavikl}
\label{lemma:typeAbddgains}
$I_{AF}(P_s)$ attains its global maximum at $\beta_{opt}=\frac{\sum_{i=1}^{M} h_{si} h_{it}}{h_{st} \sum_{i=1}^M h_{it}^2}$.
\end{pavikl}

\textit{Increasing relay power:} Let us consider the increasing total relay power scenario as in Scenario 1, Case A. Let $\beta_{max}^2 = M^u Q$. In this case, following Lemma~\ref{lemma:typeAbddgains}, we obtain the following lower bound on the achievable rate as $M \rightarrow \infty$:
\begin{eqnarray}
I_{AF}(P_s)  &>&  \frac{1}{2} \log\bigg[1 + \frac{P_s}{\sigma^2} (M h_{min} + 1)\frac{h_{s,max}^2 h_{min}}{h_{max}}\bigg], \label{eqn:upRelPwrbddgains} \\
            & &  \mbox{ if } \beta_{max} \ge \frac{\sum_{i=1}^{M} h_{si} h_{it}}{h_{st} \sum_{i=1}^M h_{it}^2}, \nonumber \\
I_{AF}(P_s)  &>&  \frac{1}{2} \log\bigg[1 + \frac{P_s}{\sigma^2} (M h_{min} + \frac{1}{\beta_{max}})\frac{h_{s,max}^2 h_{min}}{h_{max}}\bigg], \nonumber \\
            & &  \mbox{ otherwise}, \nonumber
\end{eqnarray}
where $h_{s,max} = \max\{h_{st}, h_{s1}, \ldots, h_{sM}\}$, $h_{min} = \min\{h_{1t}, \ldots, h_{Mt}\}$, and $h_{max} = \max\{h_{1t}, \ldots, h_{Mt}\}$.

\subsection{Type-B Relay Network}
\label{subsec:typeB}
Let us consider the Type-B relay network as in Figure~\ref{fig:relayNwType2}. For such networks, we consider the no-attenuation scenario where all channel gains are set to unity, that is, $h_{st} = h_{si} = h_{it} = 1$, $1 \le i \le M$ as well as $h_{i, i+1} = 1$, $1 \le i \le M-1$. The problem P4 in this case is:
\begin{pavikp}
\label{prop:typeBnoattenuation}
\begin{eqnarray}
I_{AF}(P_s, \beta) &=& \frac{1}{2} \log\bigg[1 + \frac{P_s}{\sigma^2} \frac{\big(1 + \beta M \frac{\beta^M -1}{\beta-1} - \frac{M\beta^{M+1}}{\beta-1} + \beta^2 \frac{\beta^M -1}{(\beta-1)^2}\big)^2}{1 + \sum_{i=1}^M \beta^2\big(\frac{\beta^{M-(i-1)}-1}{\beta-1}\big)^2}\bigg] \nonumber \\
                   &=& \frac{1}{2} \log\bigg[1 + \frac{P_s}{\sigma^2} \frac{\big(1 + \beta M \frac{\beta^M -1}{\beta-1} - \frac{M\beta^{M+1}}{\beta-1} + \beta^2 \frac{\beta^M -1}{(\beta-1)^2}\big)^2}{1 + \frac{\beta^2}{(\beta-1)^2}\big[\beta^2 \frac{\beta^{2M}-1}{\beta^2-1} - 2 \beta \frac{\beta^M-1}{\beta-1} + M\big]}\bigg] \nonumber \\
I_{AF}(P_s) &=& \max_{0 \le \beta \le \beta_{max}} I_{AF}(P_s, \beta) \label{eqn:typeBnoattenuation}
\end{eqnarray}
\end{pavikp}

It can be proved that the objective function is quasiconcave, therefore a unique global maximum exists. Let $\beta$ that solves \eqref{eqn:typeBnoattenuation} be denoted as $\beta=\beta_{opt}$. However, obtaining a closed-form expression for $\beta_{opt}$ does not appear straightforward, though it can be numerically computed for any $M$.

\textit{Constant total relay power:} Let us consider the case where the sum power of relay nodes is fixed irrespective of the number of relay nodes as in Scenario 1, Case B. Let $\beta_{max}^2 = \frac{Q}{M}$. As $M \rightarrow \infty$, for sufficiently large $M$, $\beta_{max} < \beta_{opt}$. Therefore, $\beta=\beta_{max}$ maximizes the achievable rate and we have the following rate achievable asymptotically as $M \rightarrow \infty$
\begin{equation}
\label{eqn:constRelPwrType2}
I_{AF}(P_s) = \frac{1}{2} \log\bigg[1 + \frac{P_s}{\sigma^2} \frac{Q M}{1+Q}\frac{1+\sqrt{Q/M}}{1-\sqrt{Q/M}}\bigg]
\end{equation}

\section{Asymptotic Capacity}
\label{sec:capacity}
In the previous two sections, we formulated the problem of maximum achievable rate for AF relay networks and then we computed explicit lower bounds to the capacity of two specific AF relay networks in various communication scenarios. In this section, we first derive an upper bound to the capacity of the general relay networks we address in this paper as introduced in the Section~\ref{sec:sysModel}. We then discuss the asymptotic behavior of the gap between this upper bound and the lower bounds computed for two specific relay networks in the previous section.

\subsection{Upper bounds to Capacity of Relay Networks}
\label{subsec:upprBdd}
In \cite{105gastparVetterli}, an upper bound to the capacity of the relay network of Type A is computed using a weaker corollary of the cut-set theorem \cite[Theorem 15.10.1]{106coverThomas} and the capacity formula for Gaussian vector channels with fixed transfer function \cite{099telatar}. Using this corollary and the capacity formula, we can also compute the upper bound to the capacity of the general relay network introduced in the Section~\ref{sec:sysModel}, as stated in the following proposition.

\begin{figure}[!t]
\centering
\includegraphics[width=3.0in]{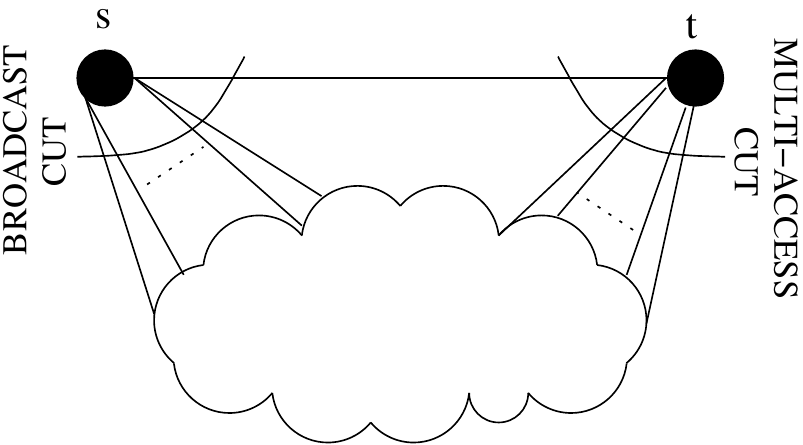}
\caption{Illustration of the particular broadcast and multiple-access cuts of the Gaussian relay network in Figure~\ref{fig:relayNw}.}
\vspace{-0.25in}
\label{fig:relayNwUprBdd}
\end{figure}

\begin{pavikp}
\label{prop:upprBdd}
The capacity $C$ of a general wireless relay network is upper-bounded as $C \le \min\{C_{BC}, C_{MAC}\}$, where $C_{BC}$ and $C_{MAC}$ are the upper bounds on the capacity of the broadcast cut and multiple-access cut respectively, as in the Figure~\ref{fig:relayNwUprBdd}, and are given as follows 
\begin{eqnarray*}
C_{BC} &=& \log\bigg[ 1 + \frac{P_s}{\sigma^2} (h_{st}^2 + \sum_{i \in S_s} h_{si}^2)\bigg] \\
C_{MAC} &=& \log\bigg[ 1 + \frac{P_s + \sum_{i \in S_t} P_i}{\sigma^2} (h_{st}^2 + \sum_{i \in S_t} h_{it}^2) \bigg]
\end{eqnarray*}
\end{pavikp}
\begin{IEEEproof}
The details of the proof are provided in the Appendix~\ref{appndx:proofsOupprBddProp}.
\end{IEEEproof}

\textit{Remark 6:} Proposition 3 in \cite{105gastparVetterli} can be obtained as a special case of the above proposition by setting $N_s = N_t = M$.

\subsection{Type A Relay Networks}
\label{subsec:typeAgaps}
\textit{No attenuation, increasing relay power:} In this case, the broadcast bound $C_{BC}$ is always asymptotically smaller than the multiple-access bound $C_{MAC}$, as follows from
\begin{equation*}
2^{C_{MAC} - C_{BC}} \approx 1 + \frac{M^{u+1} Q}{P}, \mbox{ for large } M.
\end{equation*}
Therefore it suffices to compute the asymptotic gap between $C_{BC}$ and the lower bound in \eqref{eqn:upRelPwr}. In fact, in this case we have
\begin{equation*}
C_{BC} - 2 I_{AF}(P_s) = 0, \mbox{ for all } M \ge 1
\end{equation*}
The actual capacity $C$ of the relay network in this case is bounded by $C_{BC}/2 \le C \le C_{BC}$.

\textit{No attenuation, constant total relay power:} In this case also the broadcast bound $C_{BC}$ is asymptotically smaller than the multiple access-bound $C_{MAC}$, as shown below
\begin{equation*}
\lim_{M \rightarrow \infty} 2^{C_{MAC} - C_{BC}} = 1 + \frac{Q}{P}
\end{equation*}
Therefore we only address the asymptotic gap between $C_{BC}$ and the lower bound in \eqref{eqn:constRelPwrType1}. We have
\begin{equation*}
\lim_{M \rightarrow \infty} C_{BC} - 2 I_{AF}(P_s) = \frac{1}{2} \log(1+1/Q)
\end{equation*}
The actual capacity $C$ of the relay network in this case is bounded by $\frac{1}{2}(C_{BC} - \frac{1}{2} \log(1+1/Q)) \le C \le C_{BC}$ and the bound gets tighter with increasing $Q$.

\textit{Bounded channel gains:} The gap between $C_{BC}$ and lower bound of achievable rate in \eqref{eqn:upRelPwrbddgains} is bounded asymptotically as:
\begin{equation*}
\lim_{M \rightarrow \infty} C_{BC} - 2 I_{AF}(P_s) \le \frac{1}{2}\log\bigg[\frac{h_{max}}{h_{min}^2}\bigg]
\end{equation*}

The apparent looseness of the gaps computed above compared to the corresponding gaps in \cite{105gastparVetterli} arises from the series of simplifications made to reduce the problem (P2) to the problem (P4) and the particular definition of lower bound used in \cite{105gastparVetterli}.

\subsection{Type B Relay Networks}
\label{subsec:typeBgaps}
\textit{No attenuation, constant total relay power:} As claimed above, the upper-bound in Proposition~\ref{prop:upprBdd} holds for Type B networks too. Therefore we only address the asymptotic gap between $C_{BC}$ and the lower bound in \eqref{eqn:constRelPwrType2}. We have
\begin{equation*}
\lim_{M \rightarrow \infty} C_{BC} - 2 I_{AF}(P_s) = \frac{1}{2} \log(1+1/Q)
\end{equation*}
The actual capacity $C$ of the relay network in this case too is bounded by $\frac{1}{2}(C_{BC} - \frac{1}{2} \log(1+1/Q)) \le C \le C_{BC}$ and bound gets tighter with increasing $Q$.

\section{Conclusion and Future Work}
\label{sec:conclFW}
We provide a framework to analyze the performance of the AF relay schemes in a general class of wireless relay networks. We demonstrate the effectiveness of the proposed framework in two ways. We first show that some well-established results for Gaussian relay networks (for example, those in \cite{105gastparVetterli}) can be derived in a much simpler manner. Then, we show that the achievable rates for more complex Gaussian relay networks, which cannot be addressed with existing methods, can be derived straightforwardly. The proposed framework allows for the computation of tighter amplify-and-forward lower bounds to the capacity of Gaussian relay networks.

\textit{Future Work:} We have provided the analysis of AF scheme for different scenarios with the assumption of instantaneous relay operation. However, the resulting performance is not optimal when delay in relay operation is included, as the constant power allocation is not optimal for the original problem \eqref{eqn:maxAFrateExpanded}. We also want to formally prove the hardness of optimization problem \eqref{eqn:maxAFrateExpanded} and develop efficient approximation schemes for it. An extension of our work also facilitates the computation of achievable rates for analog network coding scenarios for non-layered networks and in low to moderate SNR regimes that cannot be addressed with existing approaches. We plan to address it in detail in our future work.

\section*{Acknowldegment}
This work has been supported by the CERG grant 412207.

\appendices

\section{Proof of Lemma~\ref{lemma:infoRate}}
\label{appndx:proofInfoRate}
\setcounter{equation}{0}
\renewcommand{\theequation}{\thesection.\arabic{equation}}
Following Gallager \cite{068gallager}, we define the information rate for the channel in \eqref{eqn:ISIchnlmod} as
\begin{equation}
\label{eqn:infoRate}
I(P_s, \bm{\beta}) = \lim_{N \rightarrow \infty} I_N(P_s, \bm{\beta}),
\end{equation}
where
\begin{equation}
\label{eqn:infoRateN}
I_N(P_s, \bm{\beta}) = N^{-1} I(x_s[0,N-1]; y_t[0,N-1]),
\end{equation}
with the assumption
\begin{equation*}
x[-D^s, -1] = (0,\ldots,0)
\end{equation*}
It should be noted that as $D^s$ is finite, this assumption does not affect $I(P_s, \bm{\beta})$, but it simplifies the analysis as we show later.

Following \cite{088hirtMassey}, we introduce a new channel model whose maximum achievable rate can be easily computed. The capacity of this channel can be, subsequently, used to compute $I(P_s)$. The output of the new channel at time instant $n$ is given as
\begin{eqnarray}
\tilde{y}_t[n] &=& \sum_{j=0}^{N-1} \tilde{h}_j x_s[(n-j) \bmod N] \label{eqn:newChnl} \\
               && + \sum_{j=0}^{N-1} \tilde{h}_{1,j} z_1[(n-j) \bmod N] \nonumber \\
               && + \ldots + \sum_{j=0}^{N-1} \tilde{h}_{M,j} z_M[(n-j) \bmod N] \nonumber \\
               && + z_t[n], \quad 0 \le n < N, \nonumber
\end{eqnarray}
where $N > D^s$. Extending the unit-sample response of the channel for signal $h[0,D^S]$ with $N - D^s - 1$ zeros, we define $\tilde{h}[0,N-1] = (h_0, h_1, \ldots, h_{D^s}, 0, 0, \ldots, 0)$. We can similarly extend the corresponding unit-sample responses of the channel for noise samples $z_1, z_2, \ldots, z_M$ and define $\tilde{h}_1[0,N-1] = (h_{1,0}, h_{1,1}, \ldots, h_{1,D^1}, 0, 0, \ldots, 0)$, $\tilde{h}_2[0,N-1] = (h_{2,0}, h_{2,1}, \ldots, h_{2,D^2}, 0, 0, \ldots, 0)$, $\ldots$, $\tilde{h}_M[0,N-1] = (h_{M,0}, h_{M,1}, \ldots, h_{M,D^M}, 0, 0, \ldots, 0)$, respectively. With the definitions of these extended sample responses, we can rewrite \eqref{eqn:newChnl} symbolically as
\begin{eqnarray}
\tilde{y}_t[0,N-1] &=& x_s[0, N-1] \circledast \tilde{h}[0, N-1] \label{eqn:newChnlcircConv} \\
                   && + z_1[0, N-1] \circledast \tilde{h}_{1}[0, N-1] \nonumber \\
                   && + \ldots + z_M[0, N-1] \circledast \tilde{h}_{M}[0, N-1] \nonumber \\
                   && + z_t[0, N-1], \nonumber
\end{eqnarray}
where $ \circledast$ denotes the circular convolution operator. The transmit power constraint \eqref{eqn:pwrConstraint} leads to the power constraint 
\begin{equation}
\label{eqn:pwrConstraintnewChnl}
E[x_i^2[n]] \le P_i, \quad 0 \le n < N,
\end{equation}
for the new channel model.

The information rate for the channel defined by \eqref{eqn:newChnl} and \eqref{eqn:pwrConstraintnewChnl} is defined as
\begin{equation}
\label{eqn:infoRateNnewChnl}
\tilde{I}_N(P_s, \bm{\beta}) = N^{-1} I(x_s[0,N-1]; \tilde{y}_t[0,N-1])
\end{equation}

Taking the \textit{Discrete Fourier Transform (DFT)} on both sides of \eqref{eqn:newChnlcircConv} we have
\begin{equation}
\label{eqn:newChnlcircConvDFT}
\tilde{Y}_i = \tilde{H}_i X_i + \tilde{H}_{1,i} Z_{1,i} + \ldots + \tilde{H}_{M,i} Z_{M,i} + Z_i, \quad 0 \le i < N,
\end{equation}
where $\tilde{Y}_i$, $\tilde{H}_i$, $X_i$, $\tilde{H}_{1,i}$, $Z_{1,i}$, $\ldots$, $\tilde{H}_{M,i}$, $Z_{M,i}$, and $Z_i$ are the components of $DFT\{\tilde{y}_t[0,N-1]\}$, $DFT\{\tilde{h}[0, N-1]\}$, $DFT\{x_s[0, N-1]\}$, $DFT\{\tilde{h}_1[0, N-1]\}$, $DFT\{z_1[0, N-1]\}$, $\ldots$, $DFT\{\tilde{h}_M[0, N-1]\}$, and $DFT\{z_M[0, N-1]\}$, respectively.

Dividing both sides of this by $\tilde{H}_i$ and transforming the resulting first $\lfloor N/2 \rfloor +1$ components with the transformation \cite[eq. (24)]{088hirtMassey}, we get the following equivalent form of \eqref{eqn:newChnl} in the transform domain:
\begin{equation}
\label{eqn:newChnlTrnsfrm}
Y_i^{\prime} = X_i^{\prime} + Z_{1,i}^{\prime} + \ldots + Z_{M,i}^{\prime} + Z_i^{\prime}, \quad 0 \le i < N
\end{equation}
where $X_i^{\prime}$, $Z_{1,i}^{\prime}$, $\ldots$, $Z_{M,i}^{\prime}$, $Z_i^{\prime}$ are obtained from transform \cite[eq. (24)]{088hirtMassey} with $B_i \equiv X_i$, $B_i \equiv \tilde{H}_{1,i} Z_{1,i}/\tilde{H}_i$, $\ldots$, $B_i \equiv \tilde{H}_{M,i} Z_{M,i}/\tilde{H}_i$, and $B_i \equiv Z_i/\tilde{H}_i$, respectively.

Note that
\begin{eqnarray*}
\tilde{H}_{1,i}/\tilde{H}_i &=& \tilde{H}_{1,N-i}^{\ast}/\tilde{H}_{N-i}^{\ast}, \quad 0 \le i < N \\
&\vdots& \\
\tilde{H}_{M,i}/\tilde{H}_i &=& \tilde{H}_{M,N-i}^{\ast}/\tilde{H}_{N-i}^{\ast}, \quad 0 \le i < N \\
1/\tilde{H}_i &=& 1/\tilde{H}_{N-i}^{\ast}, \quad 0 \le i < N
\end{eqnarray*}
where $\ast$ denotes the complex conjugate. Therefore, it follows from \cite[Lemma 1]{088hirtMassey} that $Z_{m,i}^{\prime}, 1 \le m \le M$, in \eqref{eqn:newChnlTrnsfrm} are statistically independent zero mean Gaussian random variables with variance
\begin{equation}
\label{eqn:noiseVarTrnsfrm1}
\sigma_{m,i}^2 = N \sigma^2 \frac{|\tilde{H}_{m,i}|^2}{|\tilde{H}_i|^2}, \quad 0 \le i < N.
\end{equation}
Similarly, $Z_i^{\prime}$ are statistically independent zero mean Gaussian random variables with variance
\begin{equation}
\label{eqn:noiseVarTrnsfrm2}
\sigma_{i}^2 = N \sigma^2 \frac{1}{|\tilde{H}_i|^2}, \quad 0 \le i < N
\end{equation}
Therefore, the equivalent transform domain channel model \eqref{eqn:newChnlTrnsfrm} is a set of $N$ parallel discrete memoryless additive Gaussian noise channels with total noise variance
\begin{equation}
\label{eqn:noiseVarTot}
\sigma_i^2 + \sum_{m=1}^M \sigma_{m,i}^2 = N \sigma^2 \frac{1 + \sum_{m=1}^M |\tilde{H}_{m,i}|^2}{|\tilde{H}_i|^2}, \quad 0 \le i < N
\end{equation}
Further, using \cite[Lemma 2]{088hirtMassey}, it follows that the transformed inputs $X_i^{\prime}, 0 \le i, < N$, are i.i.d. zero mean Gaussian random variables with variance $N P_s$. Therefore, the equivalent transform-domain channel model in \eqref{eqn:newChnlTrnsfrm} is a set of $N$ parallel, independent, discrete, memoryless additive Gaussian noise channels with zero mean Gaussian inputs of variance $N P_s$ and total noise variance given in \eqref{eqn:noiseVarTot}. This implies that the average mutual information of $i^{\textrm{th}}$ component channel is then the capacity of point-to-point AWGN channel, given by
\begin{equation}
\label{eqn:ithComponentInfoRate}
I(X_i^{\prime}, Y_i^{\prime}) = \frac{1}{2} \log\bigg[1 + \frac{P_s}{\sigma^2} \frac{|\tilde{H}_i|^2}{1 + \sum_{m=1}^M |\tilde{H}_{m,i}|^2}\bigg]
\end{equation}
As the $N$ component channels are mutually independent, therefore $\tilde{I}_N(P_s, \bm{\beta})$ in \eqref{eqn:infoRateNnewChnl} is given by
\begin{equation}
\label{eqn:infoRateNnewChnlExpr}
\tilde{I}_N(P_s, \bm{\beta}) = \frac{1}{2N} \sum_{i=0}^{N-1} \log\bigg[1 + \frac{P_s}{\sigma^2} \frac{|\tilde{H}_i|^2}{1 + \sum_{m=1}^M |\tilde{H}_{m,i}|^2}\bigg]
\end{equation}
Further, \cite[Theorem 4]{088hirtMassey} states that
\begin{eqnarray*}
(1 - D^s/N) I_{N-D^s}(P_s, \bm{\beta}) &\le& \tilde{I}_N(P_s, \bm{\beta}) \\
                                       &\le& (1 + D^s/N) I_{N+D^s}(P_s, \bm{\beta})
\end{eqnarray*}
Therefore, we finally have $I(P_s, \bm{\beta})$ in \eqref{eqn:infoRate} as
\begin{eqnarray}
I(P_s, \bm{\beta})  &=&  \lim_{N \rightarrow \infty} \tilde{I}_N(P_s, \bm{\beta}) \label{eqn:appndxInfoRate} \\
       &=&  \lim_{N \rightarrow \infty} \frac{1}{2N} \sum_{i=0}^{N-1} \log\bigg[1 + \frac{P_s}{\sigma^2} \frac{|\tilde{H}_i|^2}{1 + \sum_{m=1}^M |\tilde{H}_{m,i}|^2}\bigg] \nonumber \\
       &\stackrel{(a)}{=}&  \frac{1}{2 \pi} \int_{-\pi}^{\pi} \frac{1}{2} \log\bigg[1 + \frac{P_s}{\sigma^2} \frac{|H(\lambda)|^2}{1 + \sum_{m=1}^M |H_{m}(\lambda)|^2}\bigg] d\lambda \nonumber \\
       &\stackrel{(b)}{=}&  \frac{1}{2 \pi} \int_{0}^{\pi} \log\bigg[1 + \frac{P_s}{\sigma^2} \frac{|H(\lambda)|^2}{1 + \sum_{m=1}^M |H_{m}(\lambda)|^2}\bigg] d\lambda \nonumber
\end{eqnarray}
where $(a)$ follows from the property of Riemann integrals as stated in \cite[Lemma 5]{088hirtMassey} and $(b)$ follows from the fact that $|H(\lambda)| = |H(-\lambda)|, |\lambda| \le \pi$. It should be noted that $H(\lambda)$ is the transfer function of the filter with unit-sample response $(h_0, h_1, \ldots, h_{D^s})$, given as follows
\begin{equation}
\label{eqn:appndxsignalTrnsFn}
H(\lambda) = \sum_{j=0}^{D^s} h_j e^{-ij\lambda}, \quad i = \sqrt{-1}
\end{equation}
Similarly, $H_m(\lambda)$ is the transfer function of the filter with unit-sample response $(h_{m,0}, h_{m,1}, \ldots, h_{m,D^m})$, $1 \le m \le M$, given by
\begin{equation}
\label{eqn:appndxrelayNoiseTrnsFn}
H_m(\lambda) = \sum_{j=0}^{D^m} h_{m,j} e^{-ij\lambda}, \quad i = \sqrt{-1}
\end{equation}

\section{Expressions for $|H(\lambda)|^2$ and $|H_m(\lambda)|^2$}
\label{appndx:explicitExprssns}
\setcounter{equation}{0}
\renewcommand{\theequation}{\thesection.\arabic{equation}}
Following \eqref{eqn:trnsFn}, we have
\begin{eqnarray}
|H(\lambda)|^2 &=& H(\lambda) H^{*}(\lambda) \nonumber \\
               &=& \sum_{j=0}^{D^s} h_j e^{-ij\lambda} \sum_{j=0}^{D^s} h_j e^{ij\lambda} \nonumber \\
               &=&  \sum_{j=0}^{D^s} h_j^2 + 2(\sum_{j=0}^{D^s-1} h_j h_{j+1}) \cos\lambda + 2 (\sum_{j=0}^{D^s-2} h_j h_{j+2}) \cos(2\lambda) \label{eqn:appndxBsig1} \\
& & + \ldots + 2 h_0 h_{D^s} \cos(D^s \lambda)  \nonumber \\
               &=& A_0 + A_{1} \cos\lambda + \ldots + A_{D^s} \cos(D^s \lambda) \label{eqn:appndxBsig2}
\end{eqnarray}
Comparing the coefficients of $\cos(j \lambda), 1 \le j \le D^s$, in \eqref{eqn:appndxBsig1} and \eqref{eqn:appndxBsig2}, defines $A_j$.

Note that we have the following relation between $A_i$s and $h_i$s:
\begin{equation}
\label{eqn:identity1}
|H(0)|^2 = \sum_{i=0}^{D^s} A_i = (h_0 + \ldots + h_{D^s})^2
\end{equation}

Noting that $\cos(n\lambda)$ can be expressed in terms of $T_n(x)$, the Chebyshev polynomials of the first kind, as $\cos(n\lambda) = T_n(\cos \lambda)$, we can rewrite $|H(\lambda)|^2$ as
\begin{eqnarray}
|H(\lambda)|^2 &=& \sum_{j=0}^{\lfloor D^s/2 \rfloor} (-1)^j A_{2j} + (\sum_{j=0}^{\lfloor D^s/2 \rfloor} (-1)^j (2j+1) A_{2j+1}) \cos\lambda + \ldots +  2^{D^s-1} A_{D^s} \cos^{D^s}\lambda \label{eqn:appndxBsig3} \\
               &=& A_0^{\prime} + A_{1}^{\prime} \cos\lambda + \ldots + A_{D^s}^{\prime} \cos^{D^s}\lambda \label{eqn:appndxBsig4}
\end{eqnarray}
Comparing the coefficients of $\cos^j\lambda, 1 \le j \le D^s$, in \eqref{eqn:appndxBsig3} and \eqref{eqn:appndxBsig4}, defines $A_j^{\prime}$.

Similarly, following \eqref{eqn:trnsFn}, the coefficients $A_{m,j}^{\prime}$ and $A_{m,j}$, $1 \le j \le D^m$, are obtained as follows:
\begin{eqnarray}
|H_m(\lambda)|^2 &=& H_m(\lambda) H_m^{*}(\lambda) \nonumber \\
               &=& \sum_{j=0}^{D^m} h_{m,j} e^{-ij\lambda} \sum_{j=0}^{D^m} h_{m,j} e^{ij\lambda} \nonumber \\
               &=& \sum_{j=0}^{D^m} h_{m,j}^2 + 2(\sum_{j=0}^{D^m-1} h_{m,j} h_{m,j+1}) \cos\lambda  + 2(\sum_{j=0}^{D^m-2} h_{m,j} h_{m,j+2}) \cos(2\lambda) \nonumber \\
               & & + \ldots + 2 h_{m,0} h_{m,D^m} \cos(D^m \lambda) \nonumber \\
               &\stackrel{(a)}{=}& A_{m,0} + A_{m,1} \cos\lambda + \ldots + A_{m,D^m} \cos(D^m \lambda) \label{eqn:appndxBnoise1} \\
               &=& \sum_{j=0}^{\lfloor D^m/2 \rfloor} (-1)^j A_{2j} + (\sum_{j=0}^{\lfloor D^m/2 \rfloor} (-1)^j (2j+1) A_{2j+1}) \cos\lambda \nonumber \\
               & & + \ldots +  2^{D^m-1} A_{D^m} \cos^{D^m}\lambda \nonumber \\
               &\stackrel{(b)}{=}& A_{m,0}^{\prime} + A_{m,1}^{\prime} \cos\lambda + \ldots + A_{m,D^m}^{\prime} \cos^{D^m}\lambda \label{eqn:appndxBnoise2}
\end{eqnarray}
The coefficients $A_{m,i}$ and $A_{m,i}^{\prime}$ are defined by comparing the cosine terms in $(a)$ and $(b)$, respectively, with the corresponding terms in the previous equations.

We have the following relation between $A_{m,i}$s and $h_{m,i}$s:
\begin{equation}
\label{eqn:identity2}
|H_m(0)|^2 = \sum_{i=0}^{D^m} A_{m,i} = (h_{m,0} + \ldots + h_{m,D^m})^2
\end{equation}

We can rewrite $\sum_{m=1}^M |H_m(\lambda)|^2$ as
\begin{eqnarray}
1 &+& \sum_{m=1}^M |H_m(\lambda)|^2 \nonumber \\
&=& (1 + \sum_{m=1}^M A_{m,0}) + (\sum_{m=1}^M A_{m,1}) \cos\lambda + \ldots + (\sum_{m=1}^M A_{m,D^n}) \cos(D^n \lambda) \nonumber \\
&\stackrel{(a)}{=}& B_0 + B_{1} \cos\lambda + \ldots + B_{D^n} \cos(D^n \lambda) \label{eqn:appndxBnoiseSum1} \\
&=& (1 + \sum_{m=1}^M A_{m,0}^{\prime}) + (\sum_{m=1}^M A_{m,1}^{\prime}) \cos\lambda + \ldots + (\sum_{m=1}^M A_{m,D^n}^{\prime}) \cos^{D^n}\lambda \nonumber \\
&\stackrel{(a)}{=}& B_0^{\prime} + B_{1}^{\prime} \cos\lambda + \ldots + B_{D^n}^{\prime} \cos^{D^n}\lambda
\end{eqnarray}
where $D^n = \max(D^1, \ldots, D^M)$. The coefficients $B_{i}$ and $B_{i}^{\prime}$ are defined by comparing the cosine terms in $(a)$ and $(b)$ with the corresponding terms in the previous equations.

\section{Illustration of the Problem P2 for Two Types of Relay Networks}
\label{appndx:probFormulnIllustrn}
\begin{figure}[!t]
\centering
\includegraphics[width=3.0in]{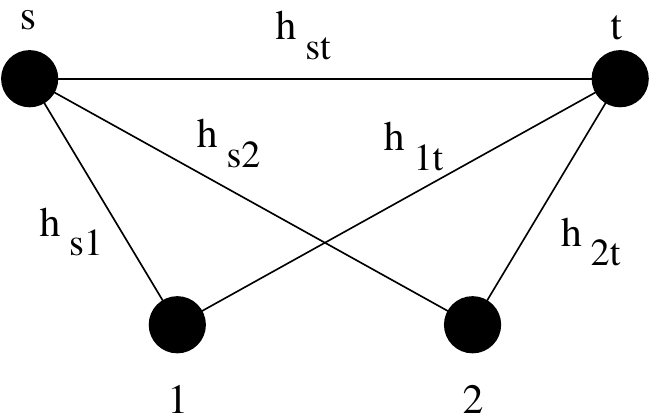}
\caption{Relay Network of Type A with M=2.}
\label{fig:configA}
\end{figure}

\setcounter{equation}{0}
\renewcommand{\theequation}{\thesection.\arabic{equation}}
\textit{Type A Relay Network:} Let us first consider the relay network configuration in Figure~\ref{fig:relayNwType1}. This is the simplest instance of the general class of relay networks we consider in this paper. For $M=2$ relay nodes, the Type A relay network is given in the Figure~\ref{fig:configA}. We illustrate the problem formulation \eqref{eqn:maxAFrateExpanded} for the network in Figure~\ref{fig:configA}.

The ISI channel between the source and the destination for this network is:
\begin{eqnarray}
y[n]  &=& h_{st} x[n] + (h_{s1} \beta_1 h_{1t} + h_{s2} \beta_2 h_{2t}) x[n-1] \label{eqn:A}\\
      &&  + \beta_1 h_{1t} z_1[n-1] + \beta_2 h_{2t} z_2[n-1] + z[n], \nonumber
\end{eqnarray}
where
\begin{eqnarray*}
0 \le \beta_1^2 \le \beta_{1, max}^2 = \frac{P_1}{h_{s1}^2 P_s + \sigma^2} \\
0 \le \beta_2^2 \le \beta_{2, max}^2 = \frac{P_2}{h_{s2}^2 P_s + \sigma^2}
\end{eqnarray*}

Comparing \eqref{eqn:A} with \eqref{eqn:ISIchnlmod} defines corresponding \textit{modified} channel parameters as follows.
\begin{eqnarray*}
h_0 &=& h_{st}, \quad h_1 = h_{s1} \beta_1 h_{1t} + h_{s2} \beta_2 h_{2t} \\
h_{1,1} &=& \beta_1 h_{1t}, \quad h_{2,1} = \beta_2 h_{2t}
\end{eqnarray*}
Using definitions in \eqref{eqn:appndxBsig2} and \eqref{eqn:appndxBnoiseSum1} results in:
\begin{eqnarray*}
A_0 &=& h_0^2 + h_1^2, \quad A_1 = 2 h_0 h_1 \\
B_0 &=& 1 + h_{1,1}^2 + h_{2,1}^2
\end{eqnarray*}
Therefore, we have the following formulation of the problem \eqref{eqn:maxAFrateExpanded} for the relay network in Figure~\ref{fig:configA}:
\begin{equation}
\label{eqn:rateA}
I_{AF}(P_s) = \max \frac{1}{2 \pi} \int_{0}^{\pi} \log \bigg[1 + \frac{P_s}{\sigma^2} \frac{A_0 + A_1 \cos{\lambda}}{B_0} \bigg] d\lambda
\end{equation}
\begin{eqnarray*}
\mbox{subject to } \hspace{1.0in} 0 &\le& \beta_1^2 \\
                                  0 &\le& \beta_2^2 \\
 \beta_1^2(h_{s1}^2 P_s + \sigma^2) &\le& P_1 \\
 \beta_2^2(h_{s2}^2 P_s + \sigma^2) &\le& P_2
\end{eqnarray*}

\begin{figure}[!t]
\centering
\includegraphics[width=3.0in]{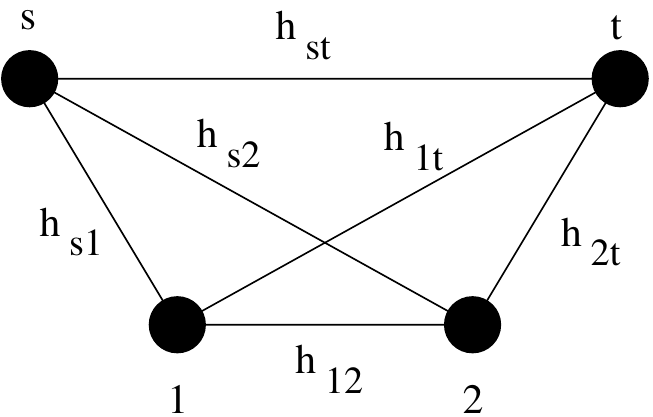}
\caption{Relay Network of Type B with M=2.}
\label{fig:configD}
\end{figure}

\textit{Type B Relay Network:} Let us next consider the network in Figure~\ref{fig:relayNwType2}. For $M=2$ relay nodes, the Type B relay network is given in the Figure~\ref{fig:configD}. For $M=2$ relay nodes, the source-destination channel for this relay configuration is:
\begin{eqnarray}
y[n]  &=&  h_{st} x[n] + + z[n] \label{eqn:D}\\
            &&  + (h_{s1} \beta_1 h_{1t} + h_{s2} \beta_2 h_{2t}) x[n-1] \nonumber \\
            &&  + (h_{s1} \beta_1 h_{12} \beta_2 h_{2t} + h_{s2} \beta_2 h_{21} \beta_1 h_{1t}) x[n-2] \nonumber \\
            &&  + \beta_1 h_{1t} z_1[n-1] + \beta_1 h_{12} \beta_2 h_{2t} z_1[n-2] \nonumber \\
            &&  + \beta_2 h_{2t} z_2[n-1] + \beta_2 h_{21} \beta_1 h_{1t} z_2[n-2], \nonumber
\end{eqnarray}
where
\begin{eqnarray*}
0 \le \beta_1^2 \le \beta_{1,max}^2  &=&  \frac{P_1}{(h_{s2}^2 \beta_2^2 h_{21}^2 + h_{s1}^2) P_s + (\beta_2^2 h_{21}^2 + 1) \sigma^2} \\
0 \le \beta_2^2 \le \beta_{2,max}^2  &=&  \frac{P_2}{(h_{s1}^2 \beta_1^2 h_{12}^2 + h_{s2}^2) P_s + (\beta_1^2 h_{12}^2 + 1) \sigma^2}
\end{eqnarray*}

Comparing \eqref{eqn:A} with \eqref{eqn:ISIchnlmod} defines corresponding \textit{modified} channel parameters as follows.
\begin{eqnarray*}
&& \hspace{-0.3in} h_0 = h_{st}, \quad h_1 = h_{s1} \beta_1 h_{1t} + h_{s2} \beta_2 h_{2t}, \quad h_2 = h_{s1} \beta_1 h_{12} \beta_2 h_{2t} + h_{s2} \beta_2 h_{21} \beta_1 h_{1t} \\
&& h_{1,1} = \beta_1 h_{1t}, \quad h_{1,2} = \beta_1 h_{12} \beta_2 h_{2t}, \quad h_{2,1} = \beta_2 h_{2t}, \quad h_{2,2} = \beta_2 h_{21} \beta_1 h_{1t}
\end{eqnarray*}
Using definitions in \eqref{eqn:appndxBsig2} and \eqref{eqn:appndxBnoiseSum1} results in:
\begin{eqnarray*}
A_0 &=& h_0^2 + h_1^2 + h_2^2, \: A_1 = 2 (h_0 + h_2) h_1, \: A_2 = 2 h_0 h_2 \\
B_0 &=& 1 + h_{1,1}^2 + h_{1,2}^2 + h_{2,1}^2 + h_{2,2}^2, \: B_1 = 2 h_{1,1} h_{1,2} + 2 h_{2,1} h_{2,2}
\end{eqnarray*}
Therefore, we have the following formulation of the problem \eqref{eqn:maxAFrateExpanded} for the relay network in Figure~\ref{fig:configD}:
\begin{equation}
\label{eqn:rateD}
\hspace{-0.025in} I_{AF}(P_s) = \max \frac{1}{2 \pi} \int_{0}^{\pi}  \log \bigg[1 + \frac{P_s}{\sigma^2} \frac{\sum_{i=0}^{2} A_i \cos{i \lambda}}{B_0 + B_1 \cos{\lambda}} \bigg] d\lambda
\end{equation}
\begin{eqnarray*}
\mbox{subject to } \hspace{2.0in} 0 &\le& \beta_1^2 \\
                                  0 &\le& \beta_2^2 \\
 \beta_1^2((h_{s2}^2 \beta_2^2 h_{21}^2 + h_{s1}^2) P_s + (\beta_2^2 h_{21}^2 + 1) \sigma^2) &\le& P_1 \\
 \beta_2^2((h_{s1}^2 \beta_1^2 h_{12}^2 + h_{s2}^2) P_s + (\beta_1^2 h_{12}^2 + 1) \sigma^2) &\le& P_2
\end{eqnarray*}
Note the nonconvex nature of the constraint set.

\section{Proofs of Lemmas and Propositions in Section~\ref{sec:computingInfoRate}}
\label{appndx:proofsOLemmasProps}
\setcounter{equation}{0}
\renewcommand{\theequation}{\thesection.\arabic{equation}}

\textit{Proof of Proposition~\ref{prop:exa1rate}:} We have from \eqref{eqn:modChnlParams} the following nonzero {modified} channel parameters:
\begin{eqnarray*}
&& h_0 = 1, \; h_1 = \sum_{j=1}^{M} \beta = M \beta \\
&& h_{m,1} = \beta, \mbox{ for all } m, 1 \le m \le M
\end{eqnarray*}
Therefore, from \eqref{eqn:identity1}, \eqref{eqn:appndxBnoise1}, and  \eqref{eqn:appndxBnoiseSum1}, we have:
\begin{eqnarray*}
A_0 + A_1 &=& (1 + M \beta)^2 \\
B_0 &=& 1 + M \beta^2
\end{eqnarray*}
Substituting these in the problem (P4) establishes the Proposition.

\textit{Proof of Lemma~\ref{lemma:typeAnoAttenuation}:} Setting the first-derivative of the objective function equal to zero gives $\beta=1$ as globally optimal solution.

Also, the objective function is quasiconcave over the domain of optimization. Note that the objective function is nondecreasing for all $\beta < 1$ and nonincreasing for all $\beta > 1$, then its quasiconcavity follows from \cite[Proposition 3]{103osborne}.

\textit{Proof of Proposition~\ref{prop:exa1ratebddgains}:} We have from \eqref{eqn:modChnlParams} the following nonzero {modified} channel parameters:
\begin{eqnarray*}
&& h_0 = h_{st}, \; h_1 = \beta \sum_{i=1}^{M} h_{si} h_{it} \\
&& h_{m,1} = \beta h_{mt}, \mbox{ for all } m, 1 \le m \le M
\end{eqnarray*}
Therefore, from \eqref{eqn:identity1}, \eqref{eqn:appndxBnoise1}, and  \eqref{eqn:appndxBnoiseSum1}, we have:
\begin{eqnarray*}
A_0 + A_1 &=& (h_{st} + \beta \sum_{i=1}^{M} h_{si} h_{it})^2 \\
B_0 &=& 1 + \beta^2 \sum_{i=1}^M h_{it}^2
\end{eqnarray*}
Substituting these in the problem (P4) establishes the Proposition.

\textit{Proof of Lemma~\ref{lemma:typeAbddgains}:} Setting the first-derivative of the objective function equal to zero gives $\beta_{opt}=\frac{\sum_{i=1}^{M} h_{si} h_{it}}{h_{st} \sum_{i=1}^M h_{it}^2}$ as the globally optimal solution.

Noting that the objective function is nondecreasing for all $\beta < \beta_{opt}$ and nonincreasing for all $\beta > \beta_{opt}$, establishes its quasiconcavity from \cite[Proposition 3]{103osborne}.

\textit{Proof of Proposition~\ref{prop:typeBnoattenuation}:} We have from \eqref{eqn:modChnlParams} the following nonzero {modified} channel parameters:
\begin{eqnarray*}
&& h_i = \sum_{j=1}^{M-(i-1)} \beta^i = (M-(i-1)) \beta^i, 1 \le i \le M,  \\
&& h_{1,i} = \beta^i, 1 \le i \le M, \\
&& h_{2,i} = \beta^1, 1 \le i \le M-1, \\
&& \vdots \\
&& h_{M,1} = \beta
\end{eqnarray*}
Therefore, from \eqref{eqn:identity1} and \eqref{eqn:identity2}, we have:
\begin{eqnarray*}
|H(0)|^2  &=&  \bigg( 1 + \beta M \frac{\beta^M -1}{\beta-1} - \frac{M\beta^{M+1}}{\beta-1} + \beta^2 \frac{\beta^M -1}{(\beta-1)^2}\bigg)^2\\
|H_m(0)|^2  &=&  \beta^2 \bigg(\frac{\beta^{M-(i-1)}-1}{\beta-1}\bigg)^2, 1 \le m \le M
\end{eqnarray*}
Substituting these expressions in \eqref{eqn:maxAFrateFnOBeta} establishes the proposition.

\section{Proof of Proposition~\ref{prop:upprBdd}}
\label{appndx:proofsOupprBddProp}
\setcounter{equation}{0}
\renewcommand{\theequation}{\thesection.\arabic{equation}}
\textbf{Broadcast upper bound:} For the broadcast cut, we have the following maximization problem:
\begin{equation*}
\max_{p_{X_s, X_1, \ldots, X_M}} I(X_s; Y_t, Y_1, \ldots, Y_M|X_1, \ldots, X_M)
\end{equation*}
subject to the power constraints
\begin{equation*}
EX_s^2 \le P_s \quad \mbox{ and } \quad EX_i^2 \le P_i, 1 \le i \le M
\end{equation*}
As the destination node knows the channel gains $h_{it}, i \in S_t$, we have
\begin{equation*}
I(X_s; Y_t, Y_1, \ldots, Y_M|X_1, \ldots, X_M) = I(X_s; \tilde{Y_t}, \tilde{Y}_{S_s})
\end{equation*}
where $\tilde{Y_t} = Y_t - \sum_{i \in S_t} h_{it} X_i$, $\tilde{Y}_{S_s} = \{\tilde{Y}_i = Y_i - \sum_{j \in {\cal N}(i) - {s}} h_{ji} X_j: i \in S_s\}$. Therefore, the maximum achievable rate of information transfer across the particular broadcast cut $C_{BC}$ of the relay network is given by
\begin{equation*}
C_{BC} = \max_{p_{X_s}: EX_s^2 \le P_s} I(X_s; \tilde{Y_t}, \tilde{Y}_{S_s})
\end{equation*}
The right-hand side can be evaluated using the capacity formula for Gaussian vector channels with fixed transfer function in \cite{099telatar}. We have a channel with scalar input and vector output, therefore corresponding channel matrix has only one singular value given by $a(N_s) = h_{st}^2 + \sum_{i \in S_s} h_{si}^2$. Hence, we have
\begin{equation}
\label{eqn:BCcap}
C_{BC} = \log\bigg[1 + \frac{P_s}{\sigma^2} a(N_s)\bigg].
\end{equation}

\textbf{Multiple-access upper bound:} For the multiple-access cut, we have the following maximization problem:
\begin{equation*}
\max_{p_{X_s, X_1, \ldots, X_M}} I(X_s, X_1, \ldots, X_M; Y_t)
\end{equation*}
subject to the power constraints
\begin{equation*}
EX_s^2 \le P_s \quad \mbox{ and } \quad EX_i^2 \le P_i, 1 \le i \le M
\end{equation*}
However, given that $Y_t = h_{st} X_s + \sum_{i \in X_{S_t}} h_{it} X_i$, we have
\begin{equation*}
I(X_s, X_1, \ldots, X_M; Y_t) = I(X_s, X_{S_t}; Y_t)
\end{equation*}
Therefore the maximization problem above reduces to
\begin{equation}
\label{eqn:MACmax}
\max_{p_{X_s, X_{S_t}}} I(X_s, X_{S_t}; Y_t)
\end{equation}
subject to the power constraints
\begin{equation}
\label{eqn:MACconstraint1}
EX_s^2 \le P_s \quad \mbox{ and } \quad EX_i^2 \le P_i, i \in S_t
\end{equation}
Let us relax this power constraint to
\begin{equation}
\label{eqn:MACconstraint2}
EX_s^2 + \sum_{i \in S_t} EX_i^2 \le P_s + \sum_{i \in S_t} P_i
\end{equation}
The solution of the problem defined by \eqref{eqn:MACmax} and \eqref{eqn:MACconstraint1} cannot be larger than the solution of the problem defined by \eqref{eqn:MACmax} and \eqref{eqn:MACconstraint2}. The latter can be evaluated by using the abovementioned result in \cite{099telatar} again. We have a channel with vector input and scalar output, therefore the corresponding channel matrix has only one singular value given by $d(N_t) = h_{st}^2 + \sum_{i \in S_t} h_{it}^2$. Therefore, the maximum achievable rate of information transfer across the particular multiple-access cut $C_{MAC}$ of the relay network is given by
\begin{equation}
\label{eqn:MACcap}
C_{MAC} = \log\bigg[ 1 + \frac{P_s + \sum_{i \in S_t} P_i}{\sigma^2} d(N_t) \bigg]
\end{equation}
As the capacity $C$ of the relay network must be smaller than the minimum of $C_{BC}$ and $C_{MAC}$, this proves the proposition.


\begin{thebibliography}{99}
\bibitem{104lanemanTseWornell} J.~N.~Laneman, D.~N.~C.~Tse, and G.~W.~Wornell, ``Cooperative diversity in wireless networks: efficient protocols and outage behavior,'' \textit{IEEE Trans. Inform. Theory,} vol. IT-50, December 2004.

\bibitem{107zhaoAdveLim} Y.~Zhao, R.~Adve, and T.~J.~Lim, ``Improving amplify-and-forward relay networks: optimal power allocation versus selection,'' \textit{IEEE Trans. Wireless. Comm.,} vol. TWC-6, August 2007.

\bibitem{107boradeZhengGallager} S.~Borade, L.~Zheng, and R.~Gallager, ``Amplify-and-forward in wireless relay networks: Rate, diversity, and network size,'' \textit{IEEE Trans. Inform. Theory,} vol. IT-53, October 2007.

\bibitem{105gastparVetterli} M.~Gastpar and M.~Vetterli, ``On the capacity of large Gaussian relay networks,'' \textit{IEEE Trans. Inform. Theory,} vol. IT-51, March 2005.

\bibitem{107gomadamJafar} K.~S.~Gomadam and S.~A.~Jafar, ``Optimal relay functionality for SNR maximization in memoryless relay networks,'' \textit{IEEE JSAC,} vol. 25, February 2007.

\bibitem{109cuiHoKliewer} T.~Cui, T.~Ho, and J.~Kliewer, ``Memoryless relay strategies for two-way relay channels,'' \textit{IEEE Trans. Comm.,} vol. 57, October 2009.

\bibitem{106zhangLiewLam} S.~Zhang, S.~C.~Liew, and P.~P.~Lam, ``Physical-Layer Network Coding,'' \textit{Proc. ACM MobiCom,} Los Angeles, CA, September 2006.

\bibitem{107kattiMaricGoldsmith} S.~Katti, I.~Mari\'{c}, A.~Goldsmith, D.~Katabi, and M.~M\'{e}dard, ``Joint relaying and network coding in wireless networks,'' \textit{Proc. IEEE ISIT 2007,} Nice, France, June 2007.

\bibitem{107kattiGollakottaKatabi} S.~Katti, S.~Gollakotta, and D.~Katabi, ``Embracing wireless interference: analog network coding,'' \textit{Proc. ACM SIGCOMM,} Kyoto, Japan, August, 2007.

\bibitem{110maricGoldsmithMedard} I.~Mari\'{c}, A.~Goldsmith, and M.~M\'{e}dard, ``Analog network coding in the high-SNR regime,'' \textit{Proc. IEEE WiNC 2010,} Boston, MA, June 2010.

\bibitem{110argyriouPandharipande} A.~Argyriou and A.~Pandharipande, ``Cooperative protocol for analog network coding in distributed wireless networks,'' \textit{IEEE Trans. Wireless. Comm.,} vol. TWC-9, October 2010.

\bibitem{111youChenLiVucetic} Q.~You, Z.~Chen, Y.~Li, and B.~Vucetic, ``Multi-hop bi-directional relay transmission schemes using amplify-and-forward and analog network coding,'' \textit{To appear in Proc. IEEE ICC 2011,} Kyoto, Japan, June 2011.

\bibitem{104zahediMohseniGamal} S.~Zahedi, M.~Mohseni, and A.~El~Gamal, ``On the capacity of AWGN relay channels with linear relaying functions,'' ,'' \textit{Proc. IEEE ISIT 2004,} Chicago, IL, June 2004.

\bibitem{103koetterMedard} R.~Koetter and M.~M\'{e}dard, ``An algebraic approach to network coding,'' \textit{IEEE/ACM Trans. Netw.,} vol. 11, October 2003.

\bibitem{088hirtMassey} W.~Hirt and J.~L.~Massey, ``Capacity of the discrete-time Gaussian channel with intersymbol interference,'' \textit{IEEE Trans. Inform. Theory,} vol. IT-34, May 1988.


\bibitem{088avrielDiewertSchaible} M.~Avriel, W.~E.~Diewert, S.~Schaible, and I.~Zang, \textit{Generalized
Concavity}. Plenum Press, 1988.

\bibitem{103schaibleShi} S.~Schaible and J.~Shi, ``Fractional Programming:
the sum-of-ratios case,'' \textit{Optimization Methods and Software,} vol. 18, April 2003.

\bibitem{107boydKimVandenberghe} S.~Boyd, S.-J.~Kim, L.~Vandenberghe, and A.~Hassibi, ``A tutorial on Geometric Programming,'' \textit{Optimization and Engineering,} vol. 8, March 2007.

\bibitem{105chiang} M.~Chiang, ``Geometric Programming for communication systems,'' Foundations and Trends in Communications and Information Theory, vol. 2, August 2005.

\bibitem{107chiangTanPalomar} M.~Chiang, C.~W.~Tan, D.~P.~Palomar, D.~O'Neill, and D.~Julian, ``Power control by Geometric Programming,'' \textit{IEEE Trans. Wireless. Comm.,} vol. TWC-6, July 2007.

\bibitem{103osborne} M.~J.~Osborne, \url{http://www.economics.utoronto.ca/osborne/MathTutorial/QCC.HTM}

\bibitem{106coverThomas} T.~M.~Cover and J.~A.~Thomas, \textit{Elements of Information Theory}. 2/e, John Wiley \& Sons, 2006.

\bibitem{099telatar} E.~Telatar, ``Capacity of multi-antenna Gaussian channels,'' \textit{Europ. Trans. Telecommun.,} vol. 10, Nov./Dec. 1999.

\bibitem{068gallager} R.~G.~Gallager, \textit{Information Theory and Reliable Communication}. John Wiley \& Sons, 1968.

\bibitem{070tsybakov} B.~S.~Tsybakov, ``Capacity of a discrete-time Gaussian channel with a filter,'' \textit{Probl. Peredach. Inform.,} vol. 6, pp. 78-82, 1970.

\bibitem{073tomsBerger} W.~Toms and T.~Berger, ``Capacity and error exponents of a channel modeled as a linear dynamic system,'' \textit{IEEE Trans. Inform. Theory,} vol. IT-19, January 1973.

\bibitem{074brandenburgWyner} L.~H. Brandenburg and A.~D.~Wyner, ``Capacity of the Guassian channel with memory: The multivariate case,'' \textit{Bell Syst. Tech. J.,} vol. 53, May-June 1974.

\bibitem{058grenanderSzego} U.~Grenander and G.~Szeg\"{o}, \textit{Toeplitz Forms and Their Applications}. 2/e, Chelsea Publishing Co., New York, 1984.

\bibitem{072gray} R.~M.~Gray, ``On the asymptotic eigenvalue distribution of Toeplitz matrices,'' \textit{IEEE Trans. Inform. Theory,} vol. IT-18, November 1972.

\bibitem{101goldsmithEffors} A.~J.~Goldsmith and M.~Effros, ``The capacity region of broadcast channels with
intersymbol interference and colored Gaussian noise,'' \textit{IEEE Trans. Inform. Theory,} vol. IT-47, January 2001.

\bibitem{109choudhuriMitra} C.~Choudhuri and U.~Mitra, ``Capacity of relay Channels with ISI and
colored Gaussian noise,'' \textit{Proc. IEEE ISIT 2009,} Seoul, Korea, June 2009.

\bibitem{111agnihotriChenJaggi} S.~Agnihotri, M.~Chen, and S.~Jaggi, ``Approximation schemes for a particular class of global optimization problems encountered in the capacity analysis of the ISI channels with colored noise,'' \textit{Under preparation,} 2011.
\end{thebibliography}
\end{document}